

%
\hyphenation{anom-aly anom-alies coun-ter-term coun-ter-terms
dif-feo-mor-phism dif-fer-en-tial super-dif-fer-en-tial dif-fer-en-tials 
super-dif-fer-en-tials par-am-e-tri-za-tion par-am-e-trized}

%
%
%
\def\unredoffs{} \def\redoffs{\voffset=-.31truein\hoffset=-.59truein}
\def\speclscape{\special{ps: landscape}}
%
%
%
%
\newbox\leftpage \newdimen\fullhsize \newdimen\hstitle \newdimen\hsbody
\tolerance=1000\hfuzz=2pt
\catcode`\@=11 
\def\bigans{b }
\message{ big or little (type b to avoid possible trouble)(b/l)? }\read-1 to\answ
\ifx\answ\bigans\message{(This will come out unreduced.}
\magnification=1200\unredoffs\baselineskip=16pt plus 2pt minus 1pt
\hsbody=\hsize \hstitle=\hsize 
\else\message{(This will be reduced.} \let\l@r=L
\magnification=1000\baselineskip=16pt plus 2pt minus 1pt \vsize=7truein
\redoffs \hstitle=8truein\hsbody=4.75truein\fullhsize=10truein\hsize=\hsbody
\output={\ifnum\pageno=0 
  \shipout\vbox{\speclscape{\hsize\fullhsize\makeheadline}
    \hbox to \fullhsize{\hfill\pagebody\hfill}}\advancepageno
  \else
  \almostshipout{\leftline{\vbox{\pagebody\makefootline}}}\advancepageno 
  \fi}
\def\almostshipout#1{\if L\l@r \count1=1 \message{[\the\count0.\the\count1]}
      \global\setbox\leftpage=#1 \global\let\l@r=R
 \else \count1=2
  \shipout\vbox{\speclscape{\hsize\fullhsize\makeheadline}
      \hbox to\fullhsize{\box\leftpage\hfil#1}}  \global\let\l@r=L\fi}
\fi
%
\newcount\yearltd\yearltd=\year\advance\yearltd by -1900

\def\Title#1#2{\nopagenumbers\abstractfont\hsize=\hstitle\rightline{#1}%
\vskip 1in\centerline{\titlefont #2}\abstractfont\vskip .5in\pageno=0}
%
%

\def\draftmode{\message{ DRAFTMODE }\def\draftdate{{\rm preliminary draft:
\number\month/\number\day/\number\yearltd\ \ \hourmin}}%
\headline={\hfil\draftdate}\writelabels\baselineskip=20pt plus 2pt minus 2pt
 {\count255=\time\divide\count255 by 60 \xdef\hourmin{\number\count255}
  \multiply\count255 by-60\advance\count255 by\time
  \xdef\hourmin{\hourmin:\ifnum\count255<10 0\fi\the\count255}}}
\def\nolabels{\def\wrlabeL##1{}\def\eqlabeL##1{}\def\reflabeL##1{}}
\def\writelabels{\def\wrlabeL##1{\leavevmode\vadjust{\rlap{\smash%
{\line{{\escapechar=` \hfill\rlap{\sevenrm\hskip.03in\string##1}}}}}}}%
\def\eqlabeL##1{{\escapechar-1\rlap{\sevenrm\hskip.05in\string##1}}}%
\def\reflabeL##1{\noexpand\llap{\noexpand\sevenrm\string\string\string##1}}}
\nolabels
%
\global\newcount\secno \global\secno=0
\global\newcount\meqno \global\meqno=1
\def\newsec#1{\global\advance\secno by1\message{(\the\secno. #1)}
\global\subsecno=0\eqnres@t\noindent{\bf\the\secno. #1}
\writetoca{{\secsym} {#1}}\par\nobreak\medskip\nobreak}
\def\eqnres@t{\xdef\secsym{\the\secno.}\global\meqno=1\bigbreak\bigskip}
\def\sequentialequations{\def\eqnres@t{\bigbreak}}\xdef\secsym{}
\global\newcount\subsecno \global\subsecno=0
\def\subsec#1{\global\advance\subsecno by1\message{(\secsym\the\subsecno. #1)}
\ifnum\lastpenalty>9000\else\bigbreak\fi
\noindent{\it\secsym\the\subsecno. #1}\writetoca{\string\quad 
{\secsym\the\subsecno.} {#1}}\par\nobreak\medskip\nobreak}
\def\appendix#1#2{\global\meqno=1\global\subsecno=0\xdef\secsym{\hbox{#1.}}
\bigbreak\bigskip\noindent{\bf Appendix #1. #2}\message{(#1. #2)}
\writetoca{Appendix {#1.} {#2}}\par\nobreak\medskip\nobreak}
%
%
\def\eqnn#1{\xdef #1{(\secsym\the\meqno)}\writedef{#1\leftbracket#1}%
\global\advance\meqno by1\wrlabeL#1}
\def\eqna#1{\xdef #1##1{\hbox{$(\secsym\the\meqno##1)$}}
\writedef{#1\numbersign1\leftbracket#1{\numbersign1}}%
\global\advance\meqno by1\wrlabeL{#1$\{\}$}}
\def\eqn#1#2{\xdef #1{(\secsym\the\meqno)}\writedef{#1\leftbracket#1}%
\global\advance\meqno by1$$#2\eqno#1\eqlabeL#1$$}
%
\newskip\footskip\footskip14pt plus 1pt minus 1pt 
\def\footnotefont{\ninepoint}\def\f@t#1{\footnotefont #1\@foot}
\def\f@@t{\baselineskip\footskip\bgroup\footnotefont\aftergroup\@foot\let\next}
\setbox\strutbox=\hbox{\vrule height9.5pt depth4.5pt width0pt}
\global\newcount\ftno \global\ftno=0
\def\foot{\global\advance\ftno by1\footnote{$^{\the\ftno}$}}
%
\newwrite\ftfile   
\def\footend{\def\foot{\global\advance\ftno by1\chardef\wfile=\ftfile
$^{\the\ftno}$\ifnum\ftno=1\immediate\openout\ftfile=foots.tmp\fi%
\immediate\write\ftfile{\noexpand\smallskip%
\noexpand\item{f\the\ftno:\ }\pctsign}\findarg}%
\def\footatend{\vfill\eject\immediate\closeout\ftfile{\parindent=20pt
\centerline{\bf Footnotes}\nobreak\bigskip\input foots.tmp }}}
\def\footatend{}
%
%
\global\newcount\refno \global\refno=1
\newwrite\rfile
\def\ref{[\the\refno]\nref}
\def\nref#1{\xdef#1{[\the\refno]}\writedef{#1\leftbracket#1}%
\ifnum\refno=1\immediate\openout\rfile=refs.tmp\fi
\global\advance\refno by1\chardef\wfile=\rfile\immediate
\write\rfile{\noexpand\item{#1\ }\reflabeL{#1\hskip.31in}\pctsign}\findarg}
\def\findarg#1#{\begingroup\obeylines\newlinechar=`\^^M\pass@rg}
{\obeylines\gdef\pass@rg#1{\writ@line\relax #1^^M\hbox{}^^M}%
\gdef\writ@line#1^^M{\expandafter\toks0\expandafter{\striprel@x #1}%
\edef\next{\the\toks0}\ifx\next\em@rk\let\next=\endgroup\else\ifx\next\empty%
\else\immediate\write\wfile{\the\toks0}\fi\let\next=\writ@line\fi\next\relax}}
\def\striprel@x#1{} \def\em@rk{\hbox{}} 
\def\lref{\begingroup\obeylines\lr@f}
\def\lr@f#1#2{\gdef#1{\ref#1{#2}}\endgroup\unskip}
\def\semi{;\hfil\break}
\def\addref#1{\immediate\write\rfile{\noexpand\item{}#1}} 
\def\listrefs{\footatend\vfill\supereject\immediate\closeout\rfile\writestoppt
\baselineskip=14pt\centerline{{\bf References}}\bigskip{\frenchspacing%
\parindent=20pt\escapechar=` \input refs.tmp\vfill\eject}\nonfrenchspacing}
\def\startrefs#1{\immediate\openout\rfile=refs.tmp\refno=#1}
\def\xref{\expandafter\xr@f}\def\xr@f[#1]{#1}
\def\refs#1{\count255=1[\r@fs #1{\hbox{}}]}
\def\r@fs#1{\ifx\und@fined#1\message{reflabel \string#1 is undefined.}%
\nref#1{need to supply reference \string#1.}\fi%
\vphantom{\hphantom{#1}}\edef\next{#1}\ifx\next\em@rk\def\next{}%
\else\ifx\next#1\ifodd\count255\relax\xref#1\count255=0\fi%
\else#1\count255=1\fi\let\next=\r@fs\fi\next}
%

%
\newwrite\ffile\global\newcount\figno \global\figno=1
\def\fig{fig.~\the\figno\nfig}
\def\nfig#1{\xdef#1{fig.~\the\figno}%
\writedef{#1\leftbracket fig.\noexpand~\the\figno}%
\ifnum\figno=1\immediate\openout\ffile=figs.tmp\fi\chardef\wfile=\ffile%
\immediate\write\ffile{\noexpand\medskip\noexpand\item{Fig.\ \the\figno. }
\reflabeL{#1\hskip.55in}\pctsign}\global\advance\figno by1\findarg}
\def\xfig{\expandafter\xf@g}\def\xf@g fig.\penalty\@M\ {}
\def\figs#1{figs.~\f@gs #1{\hbox{}}}
\def\f@gs#1{\edef\next{#1}\ifx\next\em@rk\def\next{}\else
\ifx\next#1\xfig #1\else#1\fi\let\next=\f@gs\fi\next}
\newwrite\lfile
{\escapechar-1\xdef\pctsign{\string\%}\xdef\leftbracket{\string\{}
\xdef\rightbracket{\string\}}\xdef\numbersign{\string\#}}

\def\writestop{\def\writestoppt{\immediate\write\lfile{\string\pageno%
\the\pageno\string\startrefs\leftbracket\the\refno\rightbracket%
\string\def\string\secsym\leftbracket\secsym\rightbracket%
\string\secno\the\secno\string\meqno\the\meqno}\immediate\closeout\lfile}}
\def\writestoppt{}\def\writedef#1{}
\def\seclab#1{\xdef #1{\the\secno}\writedef{#1\leftbracket#1}\wrlabeL{#1=#1}}
\def\subseclab#1{\xdef #1{\secsym\the\subsecno}%
\writedef{#1\leftbracket#1}\wrlabeL{#1=#1}}
\newwrite\tfile \def\writetoca#1{}
\def\leaderfill{\leaders\hbox to 1em{\hss.\hss}\hfill}
\def\writetoc{\immediate\openout\tfile=toc.tmp 
   \def\writetoca##1{{\edef\next{\write\tfile{\noindent ##1 
   \string\leaderfill {\noexpand\number\pageno} \par}}\next}}}

%
\catcode`\@=12 
%
\edef\tfontsize{\ifx\answ\bigans scaled\magstep3\else scaled\magstep4\fi}
\font\titlerm=cmr10 \tfontsize \font\titlerms=cmr7 \tfontsize
\font\titlermss=cmr5 \tfontsize \font\titlei=cmmi10 \tfontsize
\font\titleis=cmmi7 \tfontsize \font\titleiss=cmmi5 \tfontsize
\font\titlesy=cmsy10 \tfontsize \font\titlesys=cmsy7 \tfontsize
\font\titlesyss=cmsy5 \tfontsize \font\titleit=cmti10 \tfontsize
\skewchar\titlei='177 \skewchar\titleis='177 \skewchar\titleiss='177
\skewchar\titlesy='60 \skewchar\titlesys='60 \skewchar\titlesyss='60
\def\titlefont{\def\rm{\fam0\titlerm}
\textfont0=\titlerm \scriptfont0=\titlerms \scriptscriptfont0=\titlermss
\textfont1=\titlei \scriptfont1=\titleis \scriptscriptfont1=\titleiss
\textfont2=\titlesy \scriptfont2=\titlesys \scriptscriptfont2=\titlesyss
\textfont\itfam=\titleit \def\it{\fam\itfam\titleit}\rm}
 \ifx\answ\bigans\else scaled\magstep1\fi
\ifx\answ\bigans\def\abstractfont{\tenpoint}\else
\font\abssl=cmsl10 scaled \magstep1
\font\absrm=cmr10 scaled\magstep1 \font\absrms=cmr7 scaled\magstep1
\font\absrmss=cmr5 scaled\magstep1 \font\absi=cmmi10 scaled\magstep1
\font\absis=cmmi7 scaled\magstep1 \font\absiss=cmmi5 scaled\magstep1
\font\abssy=cmsy10 scaled\magstep1 \font\abssys=cmsy7 scaled\magstep1
\font\abssyss=cmsy5 scaled\magstep1 \font\absbf=cmbx10 scaled\magstep1
\skewchar\absi='177 \skewchar\absis='177 \skewchar\absiss='177
\skewchar\abssy='60 \skewchar\abssys='60 \skewchar\abssyss='60
\def\abstractfont{\def\rm{\fam0\absrm}
\textfont0=\absrm \scriptfont0=\absrms \scriptscriptfont0=\absrmss
\textfont1=\absi \scriptfont1=\absis \scriptscriptfont1=\absiss
\textfont2=\abssy \scriptfont2=\abssys \scriptscriptfont2=\abssyss
\textfont\itfam=\bigit \def\it{\fam\itfam\bigit}\def\footnotefont{\tenpoint}%
\textfont\slfam=\abssl \def\sl{\fam\slfam\abssl}%
\textfont\bffam=\absbf \def\bf{\fam\bffam\absbf}\rm}\fi
\def\tenpoint{\def\rm{\fam0\tenrm}
\textfont0=\tenrm \scriptfont0=\sevenrm \scriptscriptfont0=\fiverm
\textfont1=\teni  \scriptfont1=\seveni  \scriptscriptfont1=\fivei
\textfont2=\tensy \scriptfont2=\sevensy \scriptscriptfont2=\fivesy
\textfont\itfam=\tenit \def\it{\fam\itfam\tenit}\def\footnotefont{\ninepoint}%
\textfont\bffam=\tenbf \def\bf{\fam\bffam\tenbf}\def\sl{\fam\slfam\tensl}\rm}
\font\ninerm=cmr9 \font\sixrm=cmr6 \font\ninei=cmmi9 \font\sixi=cmmi6 
\font\ninesy=cmsy9 \font\sixsy=cmsy6 \font\ninebf=cmbx9 
\font\nineit=cmti9 \font\ninesl=cmsl9 \skewchar\ninei='177
\skewchar\sixi='177 \skewchar\ninesy='60 \skewchar\sixsy='60 
\def\ninepoint{\def\rm{\fam0\ninerm}
\textfont0=\ninerm \scriptfont0=\sixrm \scriptscriptfont0=\fiverm
\textfont1=\ninei \scriptfont1=\sixi \scriptscriptfont1=\fivei
\textfont2=\ninesy \scriptfont2=\sixsy \scriptscriptfont2=\fivesy
\textfont\itfam=\ninei \def\it{\fam\itfam\nineit}\def\sl{\fam\slfam\ninesl}%
\textfont\bffam=\ninebf \def\bf{\fam\bffam\ninebf}\rm} 
%
%

\hyphenation{anom-aly anom-alies coun-ter-term coun-ter-terms}
\def\inv{^{\raise.15ex\hbox{${\scriptscriptstyle -}$}\kern-.05em 1}}

\def\Dsl{\,\raise.15ex\hbox{/}\mkern-13.5mu D} 
\def\dsl{\raise.15ex\hbox{/}\kern-.57em\partial}

\font\bigit=cmti10 scaled \magstep1
\def\lspace{\ifx\answ\bigans{}\else\qquad\fi}
\def\lbspace{\ifx\answ\bigans{}\else\hskip-.2in\fi} 
\def\boxeqn#1{\vcenter{\vbox{\hrule\hbox{\vrule\kern3pt\vbox{\kern3pt
	\hbox{${\displaystyle #1}$}\kern3pt}\kern3pt\vrule}\hrule}}}
\def\mbox#1#2{\vcenter{\hrule \hbox{\vrule height#2in
		\kern#1in \vrule} \hrule}}  
%

\def\darr#1{\raise1.5ex\hbox{$\leftrightarrow$}\mkern-16.5mu #1}

\def\half{{\textstyle{1\over2}}} 
\def\roughly#1{\raise.3ex\hbox{$#1$\kern-.75em\lower1ex\hbox{$\sim$}}}
\long\def\optional#1{}
%
%
\hyphenation{reparam-etrize param-etrize reparam-etriza-tion}

\fontdimen16\tensy=2.7pt\fontdimen17\tensy=2.7pt 



%
%
%

%
%
%


\def\splitexact#1#2{\mathrel{\mathop{\null{
\lower4pt\hbox{$\rightarrow$}\atop\raise4pt\hbox{$\leftarrow$}}}\limits
^{#1}_{#2}}}
%
%

%
%
%
%

%
%

\def\ie{{\it i.e.}}

%
%


\newwrite\ffile
\def\footend{ \def\foot{\global\advance\ftno by1\chardef\wfile=\ffile
$^{\the\ftno}$\ifnum\ftno=1\immediate\openout\ffile=foots.tmp\fi%
\immediate\write\ffile{\noexpand\smallskip%
\noexpand\item{f\the\ftno:\ }\pctsign}\findarg}
\def\listrefs{\vfill\eject\immediate\closeout\ffile\parindent=20pt
\centerline{{\bf Footnotes}}\bigskip\input foots.tmp
\vfill\eject\immediate\closeout\rfile\parindent=20pt
\baselineskip14pt\centerline{{\bf References}}\bigskip\frenchspacing%
\input refs.tmp\vfill\eject\nonfrenchspacing}
} 

\global\newcount\figno \global\figno=1
\newwrite\ffile
\def\pfig#1#2{Fig.~\the\figno\nfig#1{#2}}
\def\nfig#1#2{\xdef#1{Fig. \the\figno}%
\ifnum\figno=1\immediate\openout\ffile=figs.tmp\fi%
\immediate\write\ffile{\noexpand\item{\noexpand#1\ }#2}%
\global\advance\figno by1}

%
%
\def\tfig#1{Fig.~\the\figno\xdef#1{Fig. \the\figno}\global\advance\figno by1}



\def\xx{\ $\xi$}




\def\xx{$X$}


%
%
%
%
%
%

\def~{\ifmmode\phantom{0}\else\penalty10000\ \fi}
\def\lae{\mathrel{
   \rlap{\raise 0.511ex \hbox{$<$}}{\lower 0.511ex \hbox{$\sim$}}}}
\def\gae{\mathrel{
   \rlap{\raise 0.511ex \hbox{$>$}}{\lower 0.511ex \hbox{$\sim$}}}}

\def\newpage{\vfill\eject}


\def\fun#1#2{\lower3.6pt\vbox{\baselineskip0pt\lineskip.9pt
  \ialign{$\mathsurround=0pt#1\hfil##\hfil$\crcr#2\crcr\sim\crcr}}}


%

\def\wpm{$W^{\pm}$ \thinspace}

\def\z0{$Z^0$}

\def\epem{$e^+e^-~$}
\def\epp{$e^-p~$}
\def\lplm{$l^+l^-$}
\def\wpwm{$W^+W^-~$}
\def\pt{p_T^{}}
\def\nucp{Nucl. Phys.}
\def\phr{Phys. Rev. }
\def\phrl{ Phys. Rev. Lett.}
\def\phl{ Phys. Lett. }
\Title{\vbox{\baselineskip12pt\hbox{BUHEP-93-19/hep-ph/9311351}}}
{\vbox{\centerline{The Phenomenology of a Hidden Symmetry}
	\vskip2pt\centerline{ Breaking Sector at
Electron-Positron Colliders}}}

\centerline{ M. V. Ramana$^*$}
\vskip .1in
\centerline{\it Boston University}
\centerline{\it Dept. of Physics}
\centerline{\it 590 Commonwealth Avenue}
\centerline{\it Boston, MA 02215}
\vskip .2in
\centerline{\bf ABSTRACT}

We calculate the production rate of gauge-boson pairs at \epem colliders
in a model with a ``hidden'' electroweak symmetry breaking sector - i.e.
one in which there are a large number of particles in the symmetry
breaking sector other than the $W^{\pm}$ \thinspace and the \z0. In such
a model, the elastic \wpm and \z0 scattering amplitudes are small and
structureless, i.e. lacking any discernable resonances, at all energies.
We show that two gauge boson fusion signal of electroweak symmetry
breaking is swamped by the background. Therefore, we cannot rely on
gauge boson pairs as a signal of the dynamics of symmetry breaking.
\footnote{}{$^*$ mani@budoe.bu.edu}

\newpage

The next generation of hadron, \epem and \epp colliders will search for
the electroweak symmetry breaking sector. This symmetry breaking gives
rise to Goldstone bosons, which become the longitudinally polarized
components of the \wpm and the \z0 \ref\equiv{ M.~S.~Chanowitz and
M.~K.~Gaillard, \nucp {\bf B261} (1985) {379}.}. Hence the most direct
probe of the symmetry breaking sector involves the scattering of the
longitudinally polarized $W$'s and $Z$'s.

The conventional assumption has been that there will be an enhancement
in the production of \wpwm and $Z^0Z^0$ in the form of either a weakly
interacting symmetry  breaking sector with a sub-TeV mass spectrum (~as
in the weakly coupled one doublet Higgs model~) or a strongly
interacting sector with quanta at the TeV scale (see, for example
\ref\Chan{M.~S. Chanowitz, Ann.  Rev. Nucl. Part. Sci. {\bf 38} (1988)
323 \semi M.~Golden, in {\it Beyond the Standard Model, Iowa State
University, Nov. 18-20, 1988}, K. Whisnant and B.-L. Young Eds., p. 111,
World Scientific, Singapore, 1989.}).Another possibility was pointed out
by Chivukula and Golden 
\ref\hidden{R.~S.~Chivukula and M.~Golden, \phl {\bf 267B}
(1991) 233.}.  They argued that if the electroweak symmetry breaking
sector has a large numbers of particles other than the longitudinal
components of the $W$ and $Z$, then the {\it elastic} $W$ and $Z$
scattering amplitudes can be small and structureless, i.e. lacking any
discernable resonances, at all energies.  It was further argued that in
such a model it may not be possible to rely on two gauge boson scattering
events as a signal of the symmetry breaking sector. A toy model with
these properties, based on an $O(N)$ scalar field theory solved in the
limit of large $N$, was discussed.

\nfig\feynone{Feynman diagram for Gauge Boson Fusion into Higgs resonance
and subsequent decay into a Gauge Boson Pair.}
\nfig\feyntwo{Feynman diagrams for photon photon scattering into $W^+W^-$.}

In this note, we discuss the phenomenology of this toy model at future
\epem colliders such as the NLC and the CLIC which are to operate in the
TeV range. In \epem colliders, the backgrounds to gauge boson scattering
are somewhat different from hadron colliders. If beamstrahlung and
bremstrahlung are neglected, the direct pair production of gauge bosons
is not a serious background since it can be removed by a cut on the
invariant mass. This leads us to hope that the prospects of studying
such models are better in \epem colliders. However our results show that
the ``signal'' of gauge boson fusion producing \wpwm (see 
\feynone ) will be swamped by the background from $e^+ e^- \to e^+ e^-
WW$ which proceeds through $\gamma \gamma$ scattering (see \feyntwo ).
In contrast to this, in the conventional one Higgs doublet model, the
signal is clearly visible. Thus to explore the electroweak symmetry
breaking sector in such models, the observation of other channels is
very important.  The numbers quoted and graphs shown in this note have
been computed for a \epem collider with a center of mass energy of
1~TeV. We assume a luminosity of $10^4$ inverse picobarns per year. The
results are similar for colliders with a center of mass energy of 2~TeV.

We begin by reviewing the toy model of the electroweak symmetry breaking
sector constructed in \hidden.  This model has the interesting feature
that the {\it elastic} W and Z scattering amplitudes are small and
structureless. Nevertheless the theory is strongly interacting and the
{\it total } W and Z cross sections are large - most of the cross
section is for the production of particles other than the W and the Z.

This model has two sets of fields - the $\vec{\phi}$'s and the
$\vec{\psi}$'s. The $\vec{\phi}$ and $\vec{\psi}$ are respectively $j$-
and $n$-component real scalar fields.  The Lagrangian density is
\eqn\lnought{ 
{\cal L} = \half (\partial \vec{\phi})^2 +
\half (\partial \vec{\psi})^2 - \half \mu_{0\phi}^2 \vec{\phi}^2 - \half
\mu_{0\psi}^2 \vec{\psi}^2 - {\lambda_0 \over 8 N} {(\vec{\phi}^2 +
\vec{\psi}^2)}^2.
}  
This theory has an approximate
$O(N)$ symmetry (with $N=j+n$) which is softly broken to $O(j) \times
O(n)$ so long as $\mu_{0\phi}^2
\neq \mu_{0\psi}^2$. Under the $O(j)$ and $O(n)$ symmetries the fields
$\vec{\phi}$ and $\vec{\psi}$, transform as vectors. If $\mu_{0\phi}^2$
is negative and less than $\mu_{0\psi}^2$, one of the components of
$\vec{\phi}$ gets a vacuum expectation value (VEV), breaking the
approximate $O(N)$ symmetry to $O(N-1)$. With this choice of parameters,
the exact $O(j)$ symmetry is broken to $O(j-1)$ and the theory has $j-1$
massless exact Goldstone bosons and one massive Higgs boson. The $O(n)$
symmetry is unbroken, and there are $n$ degenerate pseudo-Goldstone
bosons of mass $m_\psi$ (where $m^2_\psi =
\mu^2_{0\psi} - \mu^2_{0\phi}$). The exact Goldstone bosons will
represent the longitudinal components of the \wpm and \z0. It has been
shown that the $O(N)$ model can be solved (even for strong coupling) in
the limit of large $N$
\ref\Coleman{S.~Coleman, R.~Jackiw, and H.~D.~Politzer, \phr {\bf
D10} (1974) 2491.}. We will consider this model in the limit that $j,n
\rightarrow \infty$ with $j/n$ held fixed.

The scalar sector of the standard one-doublet Higgs model has a global $O(4)
\approx SU(2) \times SU(2)$ symmetry, where the 4 of $O(4)$ transforms as
one complex scalar doublet of the $SU(2)_W\times U(1)_Y$ electroweak
gauge interactions. It is this symmetry which is enlarged in the $O(N)$
model: the scattering amplitudes of longitudinal gauge bosons are
modelled by the corresponding $O(j)$ Goldstone boson scattering
amplitudes in the $O(j+n)$ model, solved in the large $j$ and $n$ limit.
Though $j=4$ is not particularly large, the resulting
model will have all of the correct qualitative features, the Goldstone
boson scattering amplitudes will be unitary (to the appropriate order in
$1/j$ and $1/n$), and we can investigate the theory at moderate to
strong coupling \ref\Einhorn{ M.~B.~Einhorn, \nucp {\bf B246} (1984)
75\semi R.~Casalbuoni, D.~Dominici, and R.~Gatto, \phl {\bf 147B} (1984)
419.}.  We make no assumptions about the embedding of $SU(2)_W \times
U(1)_Y$ in $O(n)$, \ie\ no assumptions about the electroweak quantum
numbers of the pseudo-Goldstone bosons \foot{ The presence of
pseudo-Goldstone bosons with electroweak or strong quantum numbers in
the theory can significantly affect the phenomenology - as an example,
gauge boson pair production at Hadron colliders in models with colored
pseudo-Goldstone bosons is discussed in detail in \ref\gscat{J.~Bagger,
S.~Dawson, and G.~Valencia, \phrl{\bf 67} (1991) {2256}\semi
R~S.~Chivukula, M.~Golden, M.~V.~Ramana, \phrl {\bf 68} (1992) {2883}
ERRATUM \phrl {\bf 69} (1992) 1291. }.}.

 Goldstone boson scattering amplitudes to leading order in $1/N$ have
been calculated in \hidden.  To this order only the
isosinglet, spin-0 amplitude $a_{00}(s)$ is nonzero:
\eqn\scatamp{
a_{00}(s)={js\over
{32\pi\left[ v^2 - Ns\left({1\over\lambda(M)}
+\widetilde{B}(s;m_\psi,M)\right)\right]}},
}
where
\def\xx{\sqrt{s / (4 m_\psi^2 - s)}}
\eqn\Btilde{
\eqalign{
\widetilde{B}(s; m_\psi, M)  = {n \over 32 N \pi^2} 
& \left\{
 1 
+ {i \over \xx} \log{i - \xx \over i + \xx}
- \log{m_\psi^2 \over M^2}
\right\}\cr
&+ 
{ j \over 32 N \pi^2 }
\left\{ 1 + \log{M^2 \over -s}
\right\}\cr
}.
}
Here $s$ is the usual Mandelstam variable, $v$ is the weak scale
(approximately 250 GeV), and $M$ is a renormalization point which acts
as a cutoff beyond which the theory is unreliable.

The amplitude $a_{00}$ of eqn.\scatamp\ can be used to derive
the ``parton level'' cross sections for $W_LW_L,Z_LZ_L \to W_LW_L~or~Z_LZ_L$. 
  This is then folded with the
appropriate gauge-boson structure functions using the ``effective W
approximation'' \ref\effectw{M.~S.~Chanowitz and M.~K.~Gaillard, \phl
{\bf 142B} (1984) {85} \semi S.~Dawson, \nucp {\bf B249}(1985){42}\semi
G.~L.~Kane, W.~W.~Repko, and W.~B.~Rolnick, \phl {\bf 148B}(1984){367}.}
to calculate the contribution of longitudinal gauge boson scattering to
the process $e^+e^- \to WW + X$. The Higgs boson appears as a resonance
in this channel. We will work at the ``parton level'' without taking
into account bremsstrahlung or beamstrahlung - including these effects
will only reduce the signal to background ratio.

If bremsstrahlung and beamstrahlung are neglected, then the invariant
mass of the final state particles is the beam energy. In that
approximation  the chief
physics background to the process $e^+e^- \to WW + X$ is 
$e^+ e^- \to e^+ e^- WW$ which proceeds through $\gamma
\gamma$ scattering (refer to \feyntwo)
\ref\gammasc{K.~Hikasa, \phl {\bf 164B}(1985){385}.} \ref\bento{
M.~C.~Bento, and C.~H.~Llewellyn Smith \nucp {\bf B289} (1987) {36}.}.
This channel, unlike the scattering of gauge bosons, does not couple to
the Higgs resonance.  Although this process is of higher order in
$\alpha$ than the fusion process, it is favoured by a large logarithmic
factor $(log({s\over{m_e^2}}))^2$, as pointed out in \gammasc.

\nfig\nxxxii{Differential production cross section for $e^+e^- \to WW$ (at a
 center of mass energy of 1 TeV) as a function of invariant $W$-pair
mass for $j=4$, $n=32$, $m_\psi = 125$GeV and the renormalization point
$M=1500$ GeV. The gauge boson scattering signal is shown as the dot-dash
curve. The background from $e^+e^- \to e^+ e^- WW$ is shown as the solid
curve. A central angle cut of $|z| < 0.7$, where $z$ is $cos(\theta)$
has been imposed on the final state $W$s.}

\nfig\nviii{Same as fig. 1 with $n=8$ and $M=2500$ GeV.}

\nfig\nz{Same as fig. 1 with $n=0$ (ref. \Einhorn) and $M=4300$ GeV.}

\nfig\higgs{Same as fig. 1 for a standard model Higgs boson with mass 485 GeV.}

We have calculated the signal and the above-mentioned background for a 
1 TeV \epem collider with the following model parameters  : 
\item {(1)} n = 32 and M = 1500 GeV. 

\item {(2)} n = 8 and M = 2500 GeV.

\item {(3)} n = 0 and M = 4300 GeV.

The first choice is inspired by the one family model of technicolor and
is one of the set of parameters studied in \hidden. The
other choices are based on \ref\nacy1 {S.~G.~Naculich and C.~P.~Yuan,
\phl {\bf B 293} (1992) {395-399}.},
where it was shown that the number of final state gauge boson pairs from
gauge boson scattering is roughly independent of $N$ (where $N = n+j$
with $j=4$) if $\sqrt{N} M$ is held fixed. The mass of the Higgs
resonance depends on the choice of $N$ and $M$ and, as is clear from
\nxxxii, \nviii ~and \nz, when the Higgs resonance becomes lighter, the
corresponding background becomes much larger and the signal is hidden.

As shown in \nxxxii ~and \nviii ~the gauge boson signal in both cases 1
and 2 are ``hidden'' because the Higgs boson is both light and broad.
Before including any branching ratios, in case 1, there are
approximately 7~events per 10~GeV bin per year at the peak while the
background is an order of magnitude higher. Thus the statistical
significance of the signal (${\rm event-rate}_{\rm signal} /
{\sqrt{{\rm event-rate}_{\rm bckgd}}}$) is less than one.  In case 2, there are
approximately 6~events per 10~GeV bin per year at the peak while the
background is about four times higher. Again the signal is significantly
below the background.  The signal in case 3 (see~\Einhorn) cannot be
detected due to the small number of events ( approx. 1 event per 10~GeV
bin per year at the peak) as can be seen in \nz.

For comparison we also calculate the signal for a standard model Higgs
boson of mass 485 GeV. This is shown in \higgs.  Since the Higgs
resonance is narrow in this case, the signal is above the background
{\it on the peak }. This corresponds to about 40 events per 10~GeV bin
per year at the peak. The background is less than 7 events per 10~GeV
bin per year in the same region. In this case the signal is fairly
visible and, as is well known, is a viable means to explore the symmetry
breaking sector. 

We have tried imposing some cuts to improve the visibility of the
signal. At higher energies, a large portion of the $e^+e^- \to e^+ e^-
WW$ background comes from the $t$ and $u$ channel $W$ exchange 
(see \feyntwo) resulting
in a large number of events in the forward/backward region.  Thus a cut
on the central angle reduces the background relative to the signal. For
example, changing from $cos(\theta) < 0.9$ to $cos(\theta) < 0.7$,
reduces the signal by a factor of about 1.3, while reducing the
background by a factor of about 3. However, since we are talking about
event rates of about 10 events per bin per year, even this is not
sufficient to improve the visibility of the signal significantly.

The background from $\gamma \gamma$ scattering (refer to \feyntwo) is not
present if the Higgs resonance is observed in the $Z^0Z^0$ channel -
provided the \z0's are unambigously identified. If the \z0's decay into
jets, then the difficulty of differentiating $Z^0$'s from \wpm is a
serious consideration.  In the neutrino mode, since both decay products
are not observed, it is not possible to reconstruct the \z0 precisely.
Only the charged lepton decay mode of the \z0 gives a clean signal. But
the branching ratio of \z0 to \lplm~is only about 3\% per leptonic
channel observed. Hence the event rate in this channel is very small and
this channel is usually not considered.  For the $n=32$ case, even if
one \z0 is taken to decay into \lplm, then the number of events is less
than 0.5~events per 10~GeV bin per year at the resonance peak.

We also do not consider other backgrounds such as the one from $e^+ e^-
\to W^+ W^- \gamma $ which, according to \ref\onephoton{G.~L.~Kane and
J.~J.~G.~Scanio, \nucp {\bf B291} (1987) {221}. }, can be reduced by
suitable cuts. The essential physics, as discussed in \onephoton, is
that energy loss from an undetected $\gamma$ is due to the escape of a
single particle, while the undetected energy in $\nu {\overline \nu} $
is carried by two particles.  Therefore the missing mass peaks at zero
for the $\gamma$, and at large masses for $\nu {\overline \nu} $. A cut
on the missing mass can clearly separate these effects. If beamstrahlung
is included, then this is not so clear. However even if this background
were significant, it would not alter any of our conclusions.

The results quoted here are calculated using both the effective W
approximation and the equivalence theorem. Since we are using these down
to fairly low energies, there would be corrections - however these
corrections are expected to be of order one
\effectw\ref\gunion{J.~F.~Gunion, J.~Kalinowski, and A. Tofighi-Niaki,
\phrl{\bf 57} (1986) {2351}.}. But the background is more than an order of
magnitude larger than the signal for the case n=32 (\nxxxii) and about
four times as large for the case n=8 (\nviii). Thus these corrections
are not expected to be significant enough to change our conclusions.  It
is clear that the signal is not detectable in the two gauge boson
channel.  To detect such a sector, one must look at other channels
besides the two gauge boson channel. In the model we have discussed, one
must look for the pseudo~Goldstone bosons - the $\psi's$ and their decay
products. The decay products would be dependent on the model. If this
were to be an extended technicolor model, then the pseudo-Goldstone
bosons are expected to decay into heavy fermions and/or jets. The
experimental signature then will be large missing transverse momentum,
of the order of the $W$ mass, and four high $\pt$ jets, from which one
can reconstruct the masses of the pseudo~Goldstone bosons ( refer
\ref\han{T. Han, ``Invited Talk at the second International Workshop on
Physics and Experiments at Linear \epem Colliders'', Waikoloa, Hawaii,
April 26-30, 1993, FERMILAB-CONF-93/217-T, July 1993, hep-ph/9307361}).
In the relatively jet-free environments of \epem colliders this could be
a spectacular signal.

In conclusion, we have shown that, in models where the symmetry breaking
sector has a large number of particles other than the \wpm and the \z0,
in experiments at the forthcoming \epem colliders operating at the TeV
range, the signal of electroweak symmetry breaking may not be 
visible in the two gauge boson channel. In such models, discovering the
electroweak breaking sector depends on the observation of the other
particles into which the \wpm and \z0 scatter into - namely the
pseudo~Goldstone bosons.

\vskip 1.5in
{\bf Acknowledgements }

I would like to thank Sekhar Chivukula for suggesting this problem and
for useful discussions. I would like to also thank Kenneth Lane, Stephen
Selipsky, Mitchell Golden, Eran Yehudai and Dimitris Kominis for useful
discussions and constructive criticism. This work was supported in part
by funds from the Texas National Research Laboratory Commission under
grant RGFY93-278, the National Science Foundation under grant PHY-9057173
and the Department of Energy under grant DE-FG02-91ER40676.

\vfill\eject\immediate\closeout\ffile
\centerline{{\bf Figure Captions}}\bigskip\frenchspacing%
\input figs.tmp\vfill\eject\nonfrenchspacing
\footatend\vfill\supereject\immediate\closeout\rfile\writestoppt
\baselineskip=14pt\centerline{{\bf References}}\bigskip{\frenchspacing%
\parindent=20pt\escapechar=` \input refs.tmp\vfill\eject}\nonfrenchspacing

\bye